\documentclass{Interspeech2024}
\usepackage{cite}
\usepackage{amsmath,amssymb,amsfonts}
\usepackage{algorithmic}
\usepackage{graphicx}
\usepackage{hyperref}
\usepackage{textcomp}
\usepackage{xcolor}
\usepackage{hyperref}
\usepackage{multirow}
\usepackage{booktabs}
\usepackage{color}
\usepackage{mathrsfs}  
\usepackage{amssymb,amsmath,bm}
\usepackage{subcaption}
\usepackage{wrapfig}
\usepackage{picinpar}
\usepackage{cutwin}
\usepackage{algorithm}  
\usepackage{algorithmic}
\usepackage{amsfonts}
\usepackage{amsfonts,amssymb}
\usepackage{url}
\usepackage{stmaryrd}

\def\BibTeX{{\rm B\kern-.05em{\sc i\kern-.025em b}\kern-.08em
    T\kern-.1667em\lower.7ex\hbox{E}\kern-.125emX}}

\interspeechcameraready

\title{Diffusion Gaussian Mixture Audio Denoise}

\name[affiliation={1}]{Pu}{Wang}
\name[affiliation={1}]{Junhui}{Li}
\name[affiliation={2}]{Jialu}{Li}
\name[affiliation={3}]{Liangdong}{Guo}
\name[affiliation={4}]{Youshan}{Zhang}

\address{
  $^1$\small Department of Mathematics, School of Science, University of Science and Technology, Liaoning, Anshan, China\\
  $^2$\small School of public policy, Cornell University, Ithaca, NY, USA\\
   $^3$\small School of Electronic and Information Engineering, University of Science and Technology Liaoning, Anshan, China\\
  $^4$\small Department of Artificial Intelligence and Computer Science,  Yeshiva University, New York, NY, US}
\email{120203803006@stu.ustl.edu.cn, Junhui\_lee@foxmail.com, jl4284@cornell.edu, Ldguo@ustl.edu.cn, youshan.zhang@yu.edu}


\begin{document}

\maketitle

\begin{abstract}

Recent diffusion models have achieved promising performances in audio-denoising tasks. The unique property of the reverse process could recover clean signals. However, the distribution of real-world noises does not comply with a single Gaussian distribution and is even unknown. The sampling of Gaussian noise conditions limits its application scenarios. To overcome these challenges, we propose a DiffGMM model, a denoising model based on the diffusion and Gaussian mixture models. We employ the reverse process to estimate parameters for the Gaussian mixture model. Given a noisy audio signal, we first apply a 1D-U-Net to extract features and train linear layers to estimate parameters for the Gaussian mixture model, and we approximate the real noise distributions. The noisy signal is continuously subtracted from the estimated noise to output clean audio signals. Extensive experimental results demonstrate that the proposed DiffGMM model achieves state-of-the-art performance.
\end{abstract}

\section{Introduction}

Audio signals are the main source of biological information transmission in nature, and various organisms interact through a wide range of sounds~\cite{li2023deeplabv3+}, such as recording recognition~\cite{7888298}, audio-to-text capabilities in social media~\cite{8424625}, and assistive hearing~\cite{schroter2020clcnet}. However, due to the existence of noise in the actual environment, the original audio becomes impure during the transmission of the audio signal. Audio denoising can significantly improve the quality of polluted audio and the accuracy of speech recognition. 
Conventional denoising methods have a significant effect on the suppression of stationary noise, but for non-stationary noise, it often cannot achieve a good noise reduction effect~\cite{zhao2018two}. Deep neural networks (DNNs) based methods commonly take a set of frequency coefficients of a short time period of the noisy signal and use paired data of noisy sounds and the corresponding clean sounds to train their denoising model~\cite{li2023dpatd,zhang2023birdsoundsdenoising}. 

Generative models include generative adversarial networks (GAN)~\cite{pascual2017segan}, variational autoencoders (VAEs)~\cite{li2021audio2gestures}, flow-based neural networks~\cite{strauss2021flow}, and diffusion models~\cite{ho2020denoising}. The diffusion model is a deep generative model that is based on two stages: a forward diffusion stage and a reverse diffusion stage. In the forward diffusion process, a Markov chain with a diffusion step (the current state is only related to the state of the previous moment) slowly adds random noise to the real data until the image becomes completely random noise. In the reverse process, data is recovered from Gaussian noise by using a series of Markov chains to gradually remove the predicted noise at each time step. In this paper, we use the reverse process from the diffusion model for audio noise reduction problem.

Diffusion probability models are a class of generation models that have shown excellent performance for image generation~\cite{nair2022image}, audio synthesis~\cite{leng2022binauralgrad}, and audio denoising~\cite{lu2021study, huang2022fastdiff}.  However, in the real condition, the distribution of noises does not comply with a single Gaussian distribution and is even unknown. One single Gaussian distribution is not enough to represent the original noise distribution. The sampling of Gaussian noise conditions limits its application scenarios. Addressing non-Gaussian noise is another challenge in diffusion models. We try to estimate the distributions of audio in the reverse process instead of the isotropic Gaussian noise. 
Gaussian mixture models can use these estimated parameters to generate approximate noise. The noisy signal is continuously subtracted from the estimated noise to output clean audio signals. In this paper, we propose a DiffGMM model, a denoising model based on the diffusion and Gaussian mixture models. Our contributions are three-fold:

\begin{itemize}
\item  We develop a diffusion Gaussian mixture model (DiffGMM), which applies the reverse process of the diffusion model to estimate parameters for the Gaussian mixture model.

\item Given a noisy audio signal, we first use a 1D-U-Net to extract features and train linear layers to estimate parameters for the Gaussian mixture model. We then approximate any arbitrary distributions of noise. By constantly subtracting the approximation noise from the original noisy audio, we can distill a clean audio signal.

\item Extensive experiments on two benchmark datasets reveal that DiffGMM outperforms state-of-the-art methods.
\end{itemize}

\begin{figure*}
	\centering
	\includegraphics[width=0.9\textwidth]{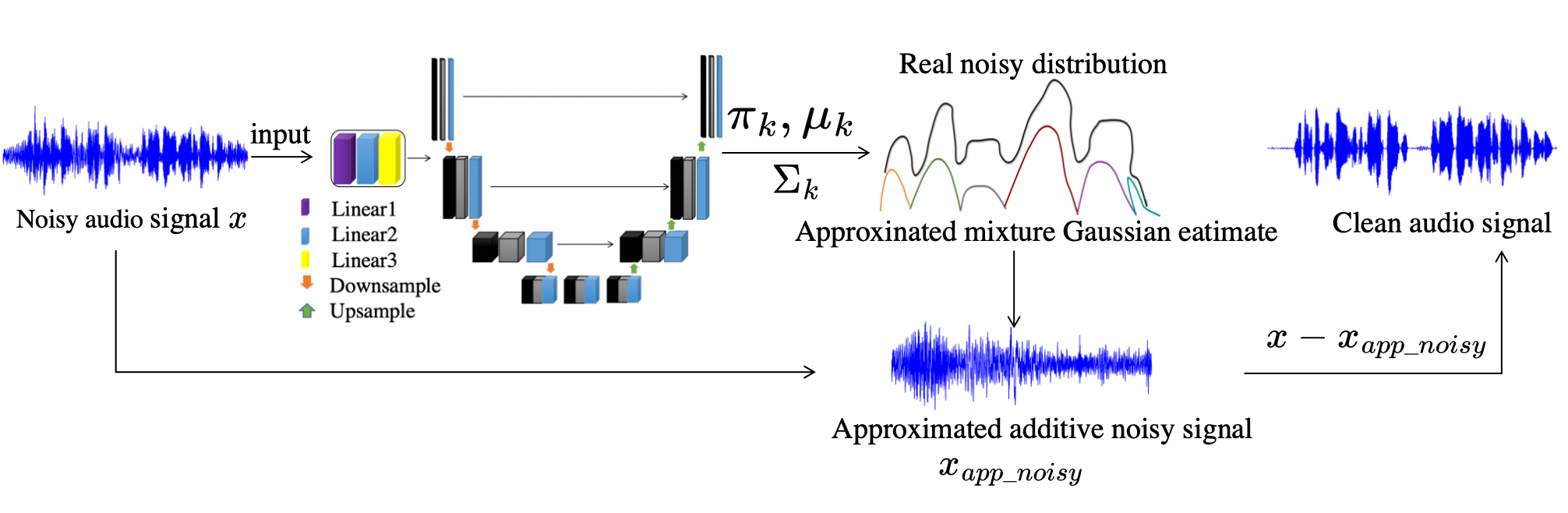}
	\caption{Flowchart of our diffusion Gaussian mixture (DiffGMM) model. We first utilize a 1D-U-Net to estimate the parameters $\pi_k, \mu_k$ and $\Sigma_k$ of GMM. We then approximate the additive noise distribution ($x_{app\_noisy}$) using GMM. The real noise is one representation to ease understanding of the GMM approximation. Finally, we continuously utilize the noisy audio signal to subtract the estimated additive noisy signal to distill a clean audio signal. }\label{Fig:arc}
 \vspace{-0.5cm}
\end{figure*}


\section{Methods}

\subsection{Problem}

A noisy audio signal $x$ can be typically expressed as:
\begin{equation}\label{eq1}
   x = y + \varepsilon  
\end{equation}
where $y$ and $\varepsilon$ denote clean audio and additive noisy signals. Given a clean audio signal and noisy audio signal $\{y_{i}\}_{i=1}^{N}$ and $\{x_{i}\}_{i=1}^{N}$, the goal of audio denoising is to extract the clean audio component $\{y_{i}\}_{i=1}^{N}$ from the noisy audio signal $\{x_{i}\}_{i=1}^{N}$ by learning a mapping $\mathcal{M}$, then minimize the approximation error between the denoised audio $\{\mathcal{M}(x_{i})\}_{i=1}^{N}$ and clean audio $\{y_{i}\}_{i=1}^{N}$. In our DiffGMM model, we use the noisy audio signal $\{x_{i}\}_{i=1}^{N}$ to continuously subtract the approximation noise to reconstruct clean signals.

\subsection{Motivation}

The diffusion model has an inherent disadvantage, $i.e.,$ a large number of sampling steps and a long sampling time because the diffusion step using Markov nuclei has only a small perturbation, but results in a large amount of diffusion. The operable model requires the same number of steps in the inference process. Therefore, it takes thousands of steps to sample the random noise until it finally changes to high-quality data similar to the prior data. At the same time, the diffusion model also limits the study of arbitrary distribution noise. Our DiffGMM model ignores the unnecessary forward process, given noisy audios are provided. We employ the reverse process to estimate parameters for the Gaussian mixture model to approximate the real noise distribution.

\subsection{Preliminary}\label{3.3}
\subsubsection{Reverse Process.} \label{3.3.2}
The reverse process is a denoising process in which $q(x_{t-1}|x_{t})$ is predicted by a neural network $p_{\theta}(x_{t-1}|x_{t})$. The reverse process converts $x_{T}$ to $x_{t}$, where $t$ represents the time $t$, and $T$ is the number of steps. We will continuously remove $T$ steps of Gaussian noise using Eq.~\eqref{eq:2}.

\begin{equation}\label{eq:2}
\setlength{\abovedisplayskip}{3pt}
   p_{\theta}(x_{0:T}):=p(x_{T})\prod^{T}_{t=1}p_{\theta}(x_{t-1}|x_{t})
\end{equation}

\begin{equation}\label{eq:9}
\setlength{\abovedisplayskip}{3pt}
  p_{\theta}(x_{t-1}|x_{t}):=N(x_{t-1}:\mu_{\theta}(x_{t},t),\Sigma_{\theta}(x_{t},t))
\end{equation}

We cannot derive $x_{t-1}$ directly from $x_{t}$ because of insufficient conditions.  We add condition $x_{0}$ to get $q(x_{t-1}|x_{t},x_{0})$, which is easy to predict $x_{t-1}$. We then get the Bayesian formula for $q(x_{t-1}|x_{t},x_{0})$ as:
\begin{equation}
  q(x_{t-1}|x_{t},x_{0})=\frac{q(x_{t}|x_{t-1},x_{0})q(x_{t-1}|x_{0})}{q(x_{t}|x_{0})}
\end{equation}

By introducing negative logarithmic likelihood $-log(p_{\theta}(x_{0}))$, we hope that the parameter $\theta$ of the neural network can make the probability of generating Eq.~\eqref{eq:2} as large as possible. However, $p_{\theta}(x_{0})$ depends on all the steps up to $x_{0}$, and $p_{\theta}(x_{0})$ is not easy to solve. The solution is to calculate the variation and lower bounds of the target:
\begin{equation*}\label{eq11}
\begin{split}
  &D_{KL}(q(x_{T}|x_{0}||p(x_{T})))\\&+ \sum^{T}_{t=2}D_{KL}(q(x_{t-1}|x_{t},x_{0})||p_{\theta}(x_{t-1}|x_{t}))
  -log(p_{\theta}(x_{0}|x_{1}))
\end{split}
\end{equation*}

From the KL divergence, we can find $q(x_{t-1}|x_{t},x_{0})$ in terms of $x_{0}$:
\begin{equation}
 q(x_{t-1}|x_{t},x_{0})=N(x_{t-1}:\widetilde{{\mu_{t}}}(x_{t},x_{0}),\widetilde{\beta}_{t}{I})
\end{equation}
where $\widetilde{\mu_{t}}(x_{t},x_{0})$ is the true value of the mean in the reverse process, $\widetilde{\beta}_{t}$ is the true value of the difference in the reverse process, $\widetilde{\beta}_{t}:=\frac{1-\widetilde{\alpha}_{t-1}}{1-\widetilde{\alpha_{t}}}{\beta_{t}}$ is fixed to a constant.

\subsubsection{Gaussian mixture model} 
Gaussian mixture model is used to combine multiple Gaussian distributions into a global distribution:
\begin{equation}\label{eq2}p(x_{i}|\theta_{k})=\sum^{K}_{k=1}\pi_{k}N(x_{i}|\mu_{k},\Sigma_{k}), i=1,...,N
\end{equation}
where, $K$ is the number of Gaussian distributions, $\theta_{k}={(\mu_{k},\Sigma_{k},\pi_{k})}$ is the collection of all unknown parameters, $\pi_{k}$ is the mixing proportions, $\mu_{k}$ is the mean vector and $\Sigma_{k}$ is the covariance matrix. The mixture coefficient $\pi_{k}$ satisfies:
 $\sum^{K}_{k=1}\pi_{k}=1, 0\leq\pi_{k}\leq1$.
Therefore, we aim to estimate the parameters $\pi_{k}, \mu_{k}, \Sigma_{k}$ for GMM in our model.

\subsection{Methodology}\label{3.4}

We assume different signals have independent and different distributions. By maximizing the product of probability density functions for all samples, we can optimize Linear layers to estimate GMM parameters $(\mu_{k},\Sigma_{k}$, and $\pi_{k})$. In other words, maximizing the product of the probability density functions of all samples is equivalent to maximizing the sum of the logarithmic probability density functions of all samples.
Given $N$ observation $\{x_n\}_{n=1}^{N}$, we take advantages of the logarithm and convert multiplication to addition. The log-likelihood function is:
\begin{equation}\label{eq4}
ln \ q(X;\pi_{1:K},\mu_{1:K},\Sigma_{1:K})=\sum^{N}_{n=1}ln(p(x_{n}|\theta_{k})).
\end{equation}

Considering arbitrary distribution $q(z)$ over the latent variables, the following decomposition always holds:
\begin{equation}\label{eq7}
	lnp(x|\theta)=\pounds(q,\theta)+KL(q||p)
\end{equation}

where 
\begin{gather}\label{eq:all1}
\scalebox{0.9}{$
\begin{aligned}
       \pounds(q,\theta)=\sum_{z}q(z)ln{\frac{p(x,z|\theta)}{q(z)}}, \ KL(q||p)=-\sum_{z}q(z)ln{\frac{p(z|x,\theta)}{q(z)}} 
\end{aligned}$}
\end{gather}

Therefore, we could use the GMM model to estimate the arbitrary distribution of audio signals. We first created an empty estimated noisy signal with the same dimension as the noisy signal and then trained the neural network $f_\theta$ to fit the noisy signal. Our goal is to solve the minimization problem:
\begin{equation}\label{Eq:goal}
    min_{\theta}||f_{\theta}(x) - (x-y)||
\end{equation}
where $x$ is the input noisy signal, $y$ is the clean signal. $x-y$ is the true noisy signal and $f_{\theta}(x)$ is the estimated noisy signal. In each iteration $i$ of training, $f_{i}$ represents the current network. By subtracting the estimated noisy signal $f_\theta$ from input noisy signal $x$, a partially denoised signal $f_i(x)$ is generated.
\begin{equation}\label{eq:denoised_sig}
    f_{i}(x) = x - f_{\theta}(x)
\end{equation}
With the increasing number of iterations, $f_{i}$ output is more expressive, and the proportion of noisy signal in $f_{i}(x)$ is getting smaller and smaller. Each iteration consists of the following steps in Alg.~\ref{alg:1}.
\begin{algorithm}

   \caption{Gaussian mixture model parameters estimation process. $I$ is the number of iterations}
   \label{alg:1}
\begin{algorithmic}[1]
   \STATE {\bfseries Input:} original audio: $x$
   \STATE {\bfseries Output:} Gaussian parameters: $\pi_{k}, \mu_{k}, \Sigma_{k}$
   \FOR{$i =1$ {\bfseries to} $I$}
   \STATE Generate initial denoising audios  by 1-D U-Net.
   \STATE $\pi_{k}, \mu_{k}, \Sigma_{k}$ by training the linear layers   
   \STATE $f_{i}\longleftarrow{f_{i-1}}$  // pass one training iteration on $f_{i-1}$starting with $\theta = \theta_{i-1}$ obtaining $f_{i}$
    \STATE Generate denoised signal $f_{i}(x)$ by Eq.~\eqref{eq:denoised_sig}
   \STATE Minimize objective optimization function by Eq.~\eqref{Eq:goal}
   \STATE Update parameters $\pi_{k}, \mu_{k}, \Sigma_{k}$
   \ENDFOR
   
   
\end{algorithmic}
\end{algorithm}

In our proposed DiffGMM model, we take the estimated noise $p(x_{i})$ in the Gaussian mixture model as the complete Gaussian noise in the diffusion process. In the new reverse process, we apply the Gaussian Markov chain model still $q (x_ {t} | x_ {0})$. We use a 1D-U-Net to extract features and train linear layers to estimate parameters for the Gaussian mixture model. According to Eq.~\eqref{eq:9}, estimating noise $p(x_{i})$, whose variance is $\Sigma_{k}$ , starts from $x_{T}$ to predict $x_{t-1}$. We can then approximate the real noise distribution $p_{DiffGMM}(x_{t-1}|x_{t})$. The noisy signal is continuously subtracted from the estimated noise to output clean audio signals:
 $p_{DiffGMM}(x_{t-1}|x_{t})=N(x_{t}|\mu,\Sigma)-\sum^{K}_{k=1}\pi_{k}N(x_{t-1}|\mu_{k},\Sigma_{k})$.

\subsubsection{Loss Function} 
In our DiffGMM model, we use the L1 loss function to train the 1D-U-Net model as follows.
\begin{equation}
    J(\theta) = \dfrac{1}{2}\sum_{i = 1}^{m}(f_{\theta}(x_{i}) - y_{i})^{2}, \quad f(\theta) = \sum_{j= 0}^{n}\theta_{j}x_{j}
\end{equation}
where $f(x)$ is the function to be fitted, $z$ is the number of records in the training set, $j$ is the number of parameters, and $\theta$ is the parameter to be iteratively solved.

We can define the evidence lower bound (ELBO) loss as the training objective of the reverse process. Based on Eq.~\eqref{eq7}, we rewrite the optimization likelihood:
\begin{equation}\label{eq:elbo}
\begin{aligned}
    ELBO&=-E_{q}(D_{KL}(q(x_{T}|x_{0})||p_{GMM\_diff}(x_{T}|x_{i}))\\&+\sum^{T}_{t=2}D_{KL}(q(x_{t-1}|x_{t})||p_{\theta}(x_{t-1}|x_{t}))-logp_{\theta}(x_{0}|x_{1})) 
\end{aligned}
\end{equation}
The first term can be ignored because there is no parameter $\theta$ in this term. The third term 
 is a known constant term, and we need to estimate only the second term:
 \begin{equation}
	L_{t-1}=\sum^{T}_{t=2}D_{KL}(q(x_{t-1}|x_{t})||p_{\theta}(x_{t-1}|x_{t}))
\end{equation}
Therefore, we have:
\begin{equation}\label{eq26}
 \begin{aligned}
     L_{t-1}=C+E_{x_{0},x_{i},\epsilon}[\frac{1}{2\Sigma_{k}^{2}}||\frac{1}{\mu_{k}}(x_{t}(x_{0},\epsilon)-{\frac{\Sigma_{k}}{\mu_{k}}})-\mu_{\theta}(x_{t}(x_{0},\epsilon)||^{2}]
 \end{aligned}
 \end{equation}

 where $C$ is a constant independent of $\theta$. By parameterizing Eq.~\eqref{eq26} simplifies to:
 \begin{equation}\label{eq18}
 \begin{split}
 L_{simple}(\theta)&:=E_{t,x_{0},x_{i},\epsilon}[||\epsilon-\epsilon_{\theta}(\mu_{k},x_{0})+(1-\mu_{k})\epsilon),t||^{2}]
  \end{split}
 \end{equation}
 where $\epsilon$ is the noise in $x_{t}$.

 \textbf{DiffGMM overall algorithm} Considering all steps in Sec.~\ref{3.4}, the scheme of our proposed DiffGMM model is shown in Fig.~\ref{Fig:arc}, and the overall algorithm is presented in Alg.~\ref{alg:VIVAD}. In Alg.~\ref{alg:1}, we get the optimal number of mixture Gaussian models of $K$ as 5, where $x_{app\_noisy}$ denotes the estimated noisy signal.

 \begin{algorithm}[ht]
   \caption{GMM and diffusion Audio Denoising (DiffGMM)}
   \label{alg:VIVAD}
\begin{algorithmic}[1]
   \STATE {\bfseries Input:} $K = 5 (k = 1, \ldots, 5)$, $\pi_{k}, \mu_{k}, \Sigma_{k}\longleftarrow{\text{1D-U-Net}}$    
   
   \FOR{$i = 1$ {\bfseries to} $N$} 
   \STATE Approximation noise $x_{app\_noisy}\longleftarrow{GMM}$   \STATE Denoised audio signal $y_{i}\longleftarrow{x_{i} - x_{app\_noisy}}$
   \ENDFOR
   \STATE {\bfseries Output:} Denoised audio signal $y$
\end{algorithmic}
\end{algorithm}

\section{Experiments}
\subsection{Datasets}
We evaluate our model using two benchmark datasets: VoiceBank-Demand~\cite{valentini2017noisy} and BirdSoundsDenoising datasets~\cite{zhang2023birdsoundsdenoising}. \textbf{VoiceBank-Demand dataset}. In this widely used noisy speech database, 251 clean speech datasets are selected from the Voice Bank corpus, including training set 252 of 11572 utterances and a test set of 872 utterances. \noindent \textbf{BirdSoundsDenoising dataset}. This bird sounds dataset contains many natural noises, including wind, waterfall, etc. The dataset is a large-scale dataset of bird sounds collected containing 10,000/1,400/2,720 in training, validation, and testing, respectively. We also choose some commonly used metrics to evaluate the enhanced speech quality~\cite{zhang2023complex,li2023dcht}, i.e., PESQ, STOI, CSIG, CBAK, and COVL. The higher these evaluation metrics are, the better the model performs. Demo samples are available at \url{https://giffgmm.github.io.}

\begin{table}[h]
  \caption{Comparison results on the VoiceBank-DEMAND dataset. ``-" means not applicable.}
  \label{Tab:1}
  \centering
\setlength{\tabcolsep}{+0.5mm}{
  \begin{tabular}{lcllllll}
    \toprule
    Methods     & Domain & PESQ & STOI & CSIG & CBAK & COVL \\
    \hline
DiffuSE(Base)~\cite{lu2021study} & T & 2.41 & - & 3.61 & 2.81 & 2.99 \\
CDiffuSE(Base)~\cite{lu2022conditional} & T & 2.44 & - & 3.66 & 2.83 & 3.03\\
PGGAN~\cite{li2022perception} &T &2.81 &0.944& 3.99& 3.59 &3.36 \\
DCCRGAN~\cite{huang2022dccrgan} &TF &2.82 &0.949 &4.01& 3.48 &3.40\\
PHASEN~\cite{yin2020phasen} &TF &2.99 &$-$ &4.18& 3.45& 3.50 \\
MetricGAN+~\cite{fu2021improved} &TF &3.15 &0.927& 4.14& 3.12 &3.52 \\
PFPL~\cite{yu2022dual} & T & 3.15 & 0.950 & 4.18 & 3.60 & 3.67 \\
MANNER~\cite{park2022manner} & T & 3.21 & 0.950 & 4.53 & 3.65 & 3.91 \\
TSTNN~\cite{wang2021tstnn} & T & 2.96 & 0.950 & 4.33 & 3.53 & 3.67 \\
DPT-FSNet~\cite{dang2022dpt} & TF & 3.33 & \textbf{0.960} & 4.58 & 3.72 & 4.00 \\
CMGAN~\cite{cao2022cmgan} & TF & 3.41 & \textbf{0.960} & 4.63 & 3.94 & 4.12 \\
\hline
DiffGMM & T & \textbf{3.48} & \textbf{0.960} & \textbf{4.72} & \textbf{4.12} & \textbf{4.34} \\
\hline
  \end{tabular}}
   \vspace{-.6cm}
\end{table}

 \begin{table}
    \centering
     \vspace{0.2cm}
     \caption{Ablation study results}\label{Tab:3}
    \setlength{\tabcolsep}{7mm} {\begin{tabular}{cc}\hline
        Model & PESQ\\ \hline
        GMM-only & 3.02 \\
        Diffusion-only & 2.79 \\
        Full Model & 3.48 \\ \hline
      \end{tabular}}
      \vspace{-0.5cm}
  \end{table}

\subsection{Implementation details}
\noindent
\textbf{Training:} The complete training pipeline is shown in Alg.~\ref{alg:VIVAD}. 
We implement our model with a Tesla P100 GPU to speed up the computation. We used the Adam optimizer for neural networks to update the network parameters $\pi_{k}$, $\mu_{k}$, $\Sigma_{k}$. During the training, the input data tensor is $8\times1\times5$ (k=5), and the output data tensor from the estimated GMM  model is $8\times1\times173568$. The 1D-U-Net has 6 layers, each with 60 filters. We use the Adam optimizer with a learning rate of $\beta_{1} = 0.9, \beta_{2} = 0.99, lr = 10^{-3}$, training iteration I = 5000, and samples were taken at 250 intervals over 5000 iterations. We train our model for 200 epochs (80 h) on a Tesla P100 GPU. The original 48 kHz files were downsampled to 16 kHz.

\subsection{Performance comparisons}
We observe that DiffGMM has the highest evaluation metrics scores in Tab.~\ref{Tab:1}. We can infer that the DiffGMM model is the best denoising model for the VoiceBank-Demand dataset among all twelve denoising models. We also reported the mean SDR of all bird sounds for the BirdSoundsDenoising dataset in both validation and test datasets. As shown in Tab.~\ref{Tab:2}, the SDR score of our DiffGMM model achieves the highest value. 

\begin{table}[h]
\small
\begin{center}
\captionsetup{font=small}
\caption{Results comparisons of different methods ($F1, IoU$, and $Dice$ scores are multiplied by 100. ``$-$" means not applicable. }
\label{Tab:2}
\setlength{\tabcolsep}{+0.5mm}{
\begin{tabular}{lllll|lllllllll}
\hline \label{tab:md}
 \multirow{2}{*}{Networks}
 &  \multicolumn{4}{c}{Validation} & \multicolumn{4}{c}{Test} \\
 \cmidrule{2-9}
& $F1$ & $IoU$  & $Dice$ & $SDR$ & $F1$ & $IoU$  & $Dice$ & $SDR$ \\
\hline
U$^2$-Net~\cite{qin2020u2}  &60.8 &45.2 &60.6 &7.85 & 60.2  &44.8 &59.9 & 7.70\\
MTU-NeT~\cite{wang2022mixed}  &69.1 &56.5 &69.0  &8.17 & 68.3  &55.7 & 68.3 &7.96  \\
Segmenter~\cite{strudel2021segmenter}  & 72.6  & 59.6 & 72.5 & 9.24 & 70.8 & 57.7 & 70.7 & 8.52   \\
U-Net~\cite{ronneberger2015u} &75.7 &64.3 &75.7 & 9.44 &74.4 &62.9 &74.4 & 8.92    \\
SegNet~\cite{badrinarayanan2017segnet} &77.5 &66.9 &77.5 & 9.55&76.1 &65.3 &76.2 & 9.43 \\
DVAD~\cite{zhang2023birdsoundsdenoising} & 82.6  & 73.5 & 82.6 & 10.33  & 81.6 & 72.3 & 81.6 & 9.96 \\
R-CED~\cite{park2016fully} & $-$ & $-$ & $-$ &2.38     &$-$ &$-$&$-$ & 1.93  \\
Noise2Noise~\cite{kashyap2021speech}  & $-$ & $-$ & $-$ & 2.40&$-$ &$-$&$-$ &1.96\\
TS-U-Net~\cite{moliner2022two}  & $-$ & $-$ & $-$ & 2.48&$-$ &$-$&$-$ &1.98\\
\hline
\textbf{DiffGMM} & $-$  & $-$ & $-$ &  \textbf{11.35}  & $-$ & $-$ & $-$ & \textbf{10.24} \\
\hline
\end{tabular}}
\end{center}
 \vspace{-.6cm}
\end{table}

\textbf{Ablation Studies}
Experimental results show that our method is remarkably effective. To further verify the effectiveness of our method, we performed an ablation analysis to show the importance of each component in our proposed model using the VoiceBank-DEMAND dataset. We conducted three experiments for the GMM-only model, the diffusion-only model, and the full model, respectively. Tab.~\ref{Tab:3} shows ablation results for a GMM-only model and diffusion-only model compared to the full model. Notably, the full model performance degrades significantly without the diffusion module.

\begin{figure}[h]
    \centering
     \vspace{0.1cm}
    \includegraphics[width=0.8\linewidth]{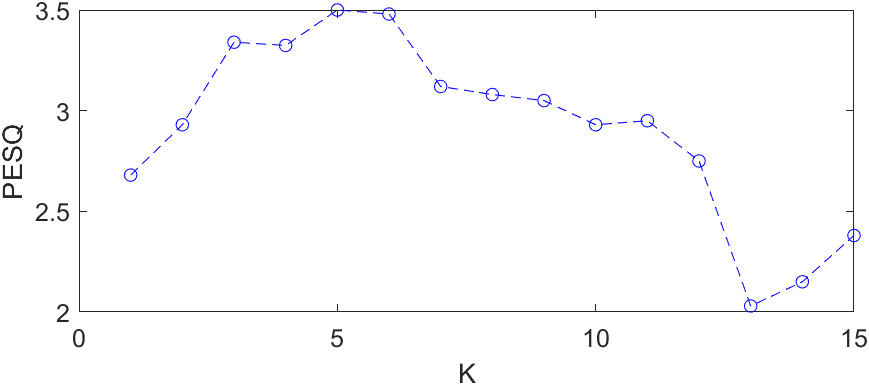}
    \caption{PESQ of different K}\label{Fig:2}
  \end{figure}

\begin{figure}[h]
	\centering
	\includegraphics[width=0.51\textwidth]{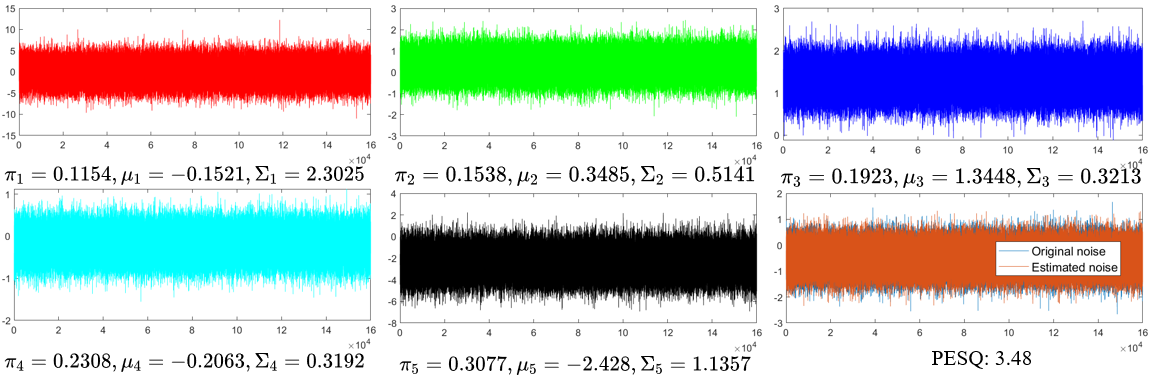}
	  \vspace{-0.5cm}
	\caption{Five different Gaussian distributions are obtained through DiffGMM in the original audio. The figure shows the Gaussian distributions corresponding to classes 1-5, and their parameters $\pi_{k}, \mu_{k}, \Sigma_{k}$ are shown below. The sixth figure is the original noisy signal and the estimated noisy signal. The X-axis is the audio length, and the Y-axis is the audio range. }\label{Fig:3}
 \vspace{-0.4cm}
\end{figure}

\textbf{Discussion}
As we can see in Fig.~\ref{Fig:2}, we show the variation curve of PESQ with different $K$. The highest PESQ = 3.48 when $K = 5$. Thus, in Fig.~\ref{Fig:3}, we show the Gaussian distributions corresponding to classes 1-5, and their parameters $\pi_{k}, \mu_{k}, \Sigma_{k}$ are shown. The sixth picture in Fig.~\ref{Fig:3} represents the original noisy signal and the estimated noisy signal. They overlap with each other, and the PESQ score is 3.48, which reflects that our proposed DiffGMM model can estimate complex noisy distributions. 

\section{Conclusion}
In this work, we develop a DiffGMM model, which is a denoising model based on the diffusion model and Gaussian mixture models. By employing the
reverse process to estimate parameters for the Gaussian mixture
model, our DiffGMM model can generalize the condition of Gaussian noise in the diffusion model to any noise distribution. Extensive experimental results demonstrate that the proposed DiffGMM model outperforms many state-of-the-art methods. 

\bibliographystyle{IEEEtran}
\bibliography{mybib}

\end{document}